\documentclass[a4paper,11pt]{article}

\usepackage{jcappub} 

\usepackage[T1]{fontenc} 

\def\bm#1{\mbox{\boldmath $#1$}}

\title{\boldmath Gravitational reheating through conformally coupled superheavy scalar particles}

\author[a,b]{Soichiro Hashiba}
\author[a,b,c]{and Jun'ichi Yokoyama}

\affiliation[a]{Department of Physics, Graduate School of Science, \\The University of Tokyo, 7-3-1 Hongo, Tokyo 113-0033, Japan}
\affiliation[b]{Research Center for the Early Universe (RESCEU), Graduate School of Science, \\The University of Tokyo, 7-3-1 Hongo, Tokyo 113-0033, Japan}
\affiliation[c]{Kavli Institute for the Physics and Mathematics of the Universe (Kavli IPMU), \\WPI, UTIAS, The University of Tokyo, 5-1-5 Kashiwanoha, Kashiwa 277-8583, Japan}

\emailAdd{sou16.hashiba@resceu.s.u-tokyo.ac.jp}
\emailAdd{yokoyama@resceu.s.u-tokyo.ac.jp}

\abstract{We calculate the number density and the energy density of a massive scalar particle conformally coupled to gravity produced by gravitational particle creation in the case the kination stage follows inflation. In this model, mode functions are derived in terms of exact or numerical solutions for the equation of motion without using the perturbative expansion. We define the adiabatic vacuum in each stage and calculate the produced number density and the energy density. The resultant power spectra show that even the superheavy particle which is as heavy as or much heavier than the Hubble scale during inflation can be produced abundantly if the transition time scale from inflation to the kination stage is smaller than the inverse of the mass of the scalar field. We also give simple forms of the reheating temperature for two cases, one that the produced scalar particles decay instantly and the other that they decay when their energy density exceeds that of inflaton.}

\begin{document}
\maketitle
\flushbottom

\section{Introduction}

During inflation, energy density of the universe is dominated by either potential energy \cite{Sato1981,Guth1981,Linde1982,Linde1983} or kinetic energy \cite{Picon1999,Kobayashi2010} of the inflaton field (see e.g. \cite{Sato2015} for a review of inflation). The $R^2$ model \cite{Starobinsky1980} may also be regarded as a potential-driven model if analyzed in the Einstein frame \cite{Maeda1988}.

Reheating or the entropy production after inflation is achieved by decay of the inflaton field oscillation \cite{Abbott1982,Dolgov1982,Traschen1990,Kofman1994,Shtanov1995,Yoshimura1995,Kofman1997} in most potential-driven models. In the case of kinetically driven inflation \cite{Picon1999,Kobayashi2010}, inflation is typically followed by a kination regime when cosmic energy density is dominated by the kinetic energy density of a free scalar field. This is also the case in some potential-driven models such as the quintessential inflation \cite{Peebles1999}.

In such models, reheating is achieved through gravitational particle production associated with the change of the cosmic expansion rate \cite{Parker1969,Zeldovich1971} from de Sitter $a(t)\propto e^{H_{\rm inf}t}$ to kination $a(t) \propto t^{1/3}$ regimes with $a(t)$ being the cosmic scale factor.

So far this problem has been analyzed by studying creation of non-conformally (often minimally) coupled massless free scalar particles \cite{Ford1987,Kunimitsu2012,Nishi2016}, since massless spinor and vector fields are conformally invariant and they are not created in this circumstances. However, such a minimally coupled massless scalar field $\phi$ acquires large quantum fluctuations during inflation. If it has a potential $V(\phi)$, its fluctuation will be saturated to drive its statistical distribution function to a static one $\rho_{\rm eq} \propto \exp \left[ -\frac{8\pi^2}{3H_{\rm inf}^4}V(\phi) \right]$, so that it behaves as a massive field with mass squared $V''(\phi)$ depending on the long-wave field value \cite{Starobinsky1994}. Thus the previous analysis \cite{Ford1987,Kunimitsu2012,Nishi2016} applies only for the case the scalar field has a shift symmetry.

Furthermore, it may be more natural to suppose the kinetic part of the scalar field is also conformally invariant with a curvature coupling $\frac{1}{12}R\phi^2$, as is often the case in models based on supergravity and superstring (see, e.g., \cite{Kachru2003,Kofman2007}).

In the present paper we study gravitational particle production of a conformally coupled massive scalar field $\phi$ as the origin of reheating after kinetically driven inflation models \cite{Picon1999,Kobayashi2010} followed by kination regime.\footnote{If there exist scalar fields with stronger non-minimal coupling, the universe may also be reheated through their spinodal instabilities, too \cite{Nakama2018,Dimopoulos2018,Fairbairn2018}, which we do not consider here.} Although the particle production rate of a massive particle has been calculated \cite{Mamaev1976,Birrell1980,Turner1988,Chung2001,Kannike2017}, almost all of them are based on the perturbative expansion such as the WKB or the Born approximation. These approximations are valid only for a small deviation from the the conformal invariance and thus it is not suitable for the calculation of a production rate of a superheavy particle. Therefore, we solve the equation of motion for mode functions without the perturbative expansion, obtain exact solutions both analytically and numerically, and calculate the number density and energy density of created particles.

This paper is organized as follows. In section \ref{sec:ac}, we analytically solve the equation of motion for a simple case where de Sitter stage is connected to kination abruptly. We derive the exact mode function in section \ref{subsec:acmf} and the adiabatic vacuum in section \ref{subsec:acadv}, and calculate the produced energy density in section \ref{subsec:acgpc}. In section \ref{sec:nc}, we numerically solve the equation of motion in a more realistic model. We construct a new model with a smooth scale factor in section \ref{subsec:ncsf}, show the result of the numerical calculation in section \ref{subsec:ncnr} and discuss its meaning in section \ref{subsec:ncd}. Our conclusion is presented in section \ref{sec:c}. We use the natural units $c = 1$, $\hbar = 1$ and $M_G = \sqrt{\hbar c/8\pi G} \approx 2.4 \times 10^{18}$ GeV throughout the paper.

\section{Analytical calculation} \label{sec:ac}
\subsection{Model and mode functions} \label{subsec:acmf}
We consider a massive scalar field $\phi(\bm{x},t)$ which is conformally coupled to gravity in a spatially flat Friedmann-Lema\^{i}tre-Robertson-Walker (FLRW) metric. The line element is
\begin{equation}
	ds^2 = a^2(\eta) (-d\eta^2 + d\bm{x}^2),
\end{equation}
where $\eta$ and $a(\eta)$ denote the conformal time and the scale factor, respectively. Here, we consider a transition from the de Sitter to the kination stage. Since the scale factor asymptotically behaves as $a(\eta)\propto-\eta^{-1}$ and $a(\eta)\propto\eta^{1/2}$ in the de Sitter and the kination stage, respectively, we connect them at the end of inflation $\eta=\eta_f$ continuously up to the first-order derivative and obtain
\begin{equation}
	a^2(\eta) = \begin{cases}
		\frac{1}{H_{\rm inf}^2 |\eta|^2} & (\eta<\eta_f<0) \\
		\frac{2}{H_{\rm inf}^2 |\eta_f|^3} (\eta-\eta_f) + \frac{1}{H_{\rm inf}^2 |\eta_f|^2} & (\eta_f<\eta)
	\end{cases},
\end{equation}
where $H_{\rm inf}$ denotes the Hubble constant during inflation. We put the scale factor at the end of inflation $a(\eta_f)$ to unity hereafter without loss of generality. Hence, the scale factor squared is given by
\begin{equation}
	a^2(\eta) = \begin{cases}
		\frac{1}{H_{\rm inf}^2 \eta^2} & (\eta<-H_{\rm inf}^{-1}) \\
		2H_{\rm inf}\eta + 3 & (\eta>-H_{\rm inf}^{-1}) \label{scale}
	\end{cases}.
\end{equation}
Such a scale factor can be realized by a cliff-shape potential. Let us assume that the potential of the inflaton is approximated by the Heaviside step function $\Theta(x)$ around the origin as $V(\varphi) \approx V_0 \Theta(-\varphi)$, where $\varphi$ is the inflaton, $V_0$ is a height of the potential plateau $V_0 = 3M_G^2 H_{\rm inf}^2$ and a slow-roll inflation is realized in $\varphi<0$. According to the equation of motion, the inflaton field develops when it falls over this precipice at $\eta = \eta_f$,
\begin{align}
	\ddot{\varphi} &\approx - \frac{dV(\varphi)}{d\varphi} \nonumber \\
	\int_{\eta_f - \epsilon}^{\eta_f + \epsilon} \ddot{\varphi}(t) \dot{\varphi}(t) dt &\approx - \int_{-\epsilon}^{+\epsilon} \frac{dV(\varphi)}{d\varphi} d\varphi \nonumber \\
	\frac{1}{2} \left[ \dot{\varphi}^2(\eta_f + \epsilon) - \dot{\varphi}^2(\eta_f - \epsilon) \right] &= V_0 \nonumber \\
	\dot{\varphi}^2 (\eta_f + \epsilon) &\approx 2V_0 \ \gg \dot{\varphi}^2 (\eta_f - \epsilon)
\end{align}
where $\epsilon$ is an infinitesimal, the first and the last line come from the slow-roll conditions and the second line from $a(\eta_f)=1$. Therefore, the inflaton field is abruptly accelerated and inflation immediately changes into kination. Since the total energy of the universe is conserved, the Hubble parameter does not change at the moment of this transition, in other words, the scale factor is continuous up to the first-order derivative. Since a crease of the inflaton potential produces a hump in the spectrum of the primordial fluctuation \cite{Starobinsky1992}, the shape of the cliff is constrained by the overproduction problem of primordial black holes. Hence, this model should be regarded as a toy model.

The Lagrangian for a conformally coupled massive scalar field is
\begin{equation}
	\mathcal{L}_\phi = \sqrt{-g} \left( -\frac{1}{2}g^{\mu\nu}\nabla_\mu \phi \nabla_\nu \phi - \frac{1}{12}\phi^2 R - \frac{1}{2}m^2 \phi^2 \right), \label{lagrangian}
\end{equation}
where $R$ and $m$ denote the scalar curvature and a mass of the scalar field, respectively. Varying (\ref{lagrangian}) with respect to $\phi$, we obtain the equation of motion (EOM) for the scalar field as
\begin{equation}
	\left( -\Box + m^2 + \frac{1}{6}R \right) \phi = 0. \label{eom1}
\end{equation}
Substituting the mode expansion of the scalar field
\begin{equation}
	\hat{\phi}(\bm{x},\eta) = \int \frac{d^3 k}{(2\pi)^{3/2}a(\eta)} \left( e^{i\bm{k}\cdot\bm{x}} \hat{a}_k \chi^\ast_k(\eta) + e^{-i\bm{k}\cdot\bm{x}} \hat{a}_k^\dag \chi_k(\eta) \right)
\end{equation}
into (\ref{eom1}) and we obtain the equation of motion for the conformally rescaled mode function $\chi_k(\eta)$ as
\begin{equation}
    	\frac{d^2 \chi_k(\eta)}{d\eta^2} + \left[ k^2 + a^2(\eta)m^2 \right] \chi_k(\eta) = 0. \label{eom}
\end{equation}

\subsection{Adiabatic vacuum} \label{subsec:acadv}
We can obtain the particle production rate by solving (\ref{eom}) and comparing its solution with the mode function of the vacuum state. However, there is no unique definition of the vacuum state in curved spacetime and hence we have to define in an appropriate way. We adopt so-called the adiabatic vacuum \cite{qeg}. Its basic idea is as follows. If the metric of a spacetime changes so slowly or equivalently, the momentum of a particle is so large, it hardly feels the expansion of the spacetime. Therefore, mode functions of particles in this adiabatic limit should asymptotically approach those in the Minkowski space. The adiabatic vacuum is the vacuum state corresponding to the mode function which asymptotically behaves as a positive frequency mode in the Minkowski space.

First, let us derive the adiabatic vacuum in the de Sitter space. Substituting (\ref{scale}) into (\ref{eom}) provided that $\eta<-H_{\rm inf}^{-1}$, we obtain the EOM in this case as
\begin{equation}
	\frac{d^2 \chi_k}{d\eta^2} + \left( k^2 + \frac{m^2}{H_{\rm inf}^2 \eta^2} \right)\chi_k = 0. \label{eomds}
\end{equation}
We can easily solve (\ref{eomds}) on the initial condition
\begin{equation}
	\chi_k(\eta) \xrightarrow{\eta\to-\infty} \frac{1}{\sqrt{2k}} e^{-ik\eta},
\end{equation}
and then obtain the adiabatic vacuum in the de Sitter space
\begin{equation}
	\chi_k^{\rm BD}(\eta) = \frac{\sqrt{-\pi\eta}}{2} e^{-i\frac{2\nu+1}{4}\pi} H^{(1)}_\nu(-k\eta), \label{advds}
\end{equation}
where $\nu = \frac{1}{2}\sqrt{1-4m^2/H_{\rm inf}^2}$. This is called the Bunch-Davies vacuum \cite{bdv}. Originally, the factor $e^{-i\frac{2\nu+1}{4}\pi}$ was not included since they considered a light particle such that $\nu\in\mathbb{R}$ and then this factor is a physically meaningless phase factor. In contrast, in our case, $\nu$ can be imaginary and this factor becomes important. We cannot satisfy the normalization condition $W[\chi_k,\chi_k^\ast] = 2i$ without this factor. Here, $W$ is the Wronskian with respect to $\eta$ defined by $W[f(\eta),g(\eta)]=fg'-f'g$. However, $e^{-i\frac{2\nu+1}{4}\pi}$ is cancelled out by this normalization and it has no effect on gravitational particle creation.

Next, the adiabatic vacuum in the kination stage. Substituting (\ref{scale}) into (\ref{eom}) with $\eta>-H_{\rm inf}^{-1}$, we obtain the EOM in this case as
\begin{equation}
	\frac{d^2 \chi_k}{d\eta^2} + \left[ k^2 + m^2(2H_{\rm inf}\eta + 3) \right] \chi_k = 0. \label{eomk}
\end{equation}
The general solutions of (\ref{eomk}) is
\begin{equation}
	\chi_k(\eta) = C_1 \sqrt{x} H^{(1)}_{1/3}\left(\frac{2}{3}x^{3/2}\right) + C_2 \sqrt{x} H^{(2)}_{1/3}\left(\frac{2}{3}x^{3/2}\right), \label{gensolnk}
\end{equation}
where $C_1$ and $C_2$ are constants independent of $\eta$, and
\begin{equation}
	x(k,\eta) = \frac{k^2 + 3m^2 + 2m^2H_{\rm inf}\eta}{(2m^2H_{\rm inf})^{2/3}}.
\end{equation}
We take the adiabatic limit of (\ref{gensolnk}) as $k^2/a^2m^2\to\infty$, and then obtain
\begin{align}
	\chi_k(\eta) &\sim C_2 \left[ \frac{k}{(2m^2H_{\rm inf})^{1/3}} + \mathcal{O} \left( \left(\frac{m^4}{H_{\rm inf}k^3}\right)^{1/3} \right) \right] \nonumber \\
	&\quad \times H^{(2)}_{1/3}\left( \frac{k^3}{3m^2H_{\rm inf}} + \frac{3k}{2H_{\rm inf}} + k\eta + \mathcal{O}\left( \frac{m^2}{H_{\rm inf}k} \right) \right) + \left( H^{(1)}_{1/3} {\rm \ term} \right) \nonumber \\
	&\sim C_2 (2m^2H_{\rm inf})^{1/6} \sqrt{\frac{3}{\pi k}} \exp\left[ - \left( \frac{k^3}{3m^2H_{\rm inf}} + \frac{3k}{2H_{\rm inf}} + k\eta - \frac{5}{12}\pi \right) i \right] \nonumber \\
	&\quad + C_1 (2m^2H_{\rm inf})^{1/6} \sqrt{\frac{3}{\pi k}} \exp\left[ \left( \frac{k^3}{3m^2H_{\rm inf}} + \frac{3k}{2H_{\rm inf}} + k\eta - \frac{5}{12}\pi \right) i \right] + \mathcal{O}\left(\left( \frac{m^2H_{\rm inf}}{k^3} \right)^{7/6} \right).
\end{align}
Hence, we take
\begin{equation}
	C_1 = 0, \qquad C_2 = \sqrt{\frac{\pi}{6}} \:(2m^2H_{\rm inf})^{-1/6} \exp\left[ \left( \frac{k^3}{3m^2H_{\rm inf}} + \frac{3k}{2H_{\rm inf}} -\frac{5}{12}\pi \right) i \right]
\end{equation}
so that (\ref{gensolnk}) satisfies the adiabatic condition
\begin{equation}
	\chi_k(\eta) \xrightarrow{\frac{k^2}{a^2m^2}\to\infty} \frac{1}{\sqrt{2k}} e^{-ik\eta}.
\end{equation}
Thus, we obtain the adiabatic vacuum in the kinetic energy dominated flat universe as
\begin{equation}
	\chi_k^{\rm K}(\eta) = \sqrt{\frac{\pi}{6}} \:(2m^2H_{\rm inf})^{-1/6} \exp\left[ \left( \frac{k^3}{3m^2H_{\rm inf}} + \frac{3k}{2H_{\rm inf}} -\frac{5}{12}\pi \right) i \right] \sqrt{x} H^{(2)}_{1/3}\left(\frac{2}{3}x^{3/2}\right). \label{advk}
\end{equation}
In this paper, we call this ``k-vacuum.'' This can also be written in terms of Airy functions \cite{Lankinen2017}.

Assuming that the universe is in the Bunch-Davies vacuum during inflation, then the solution of (\ref{eom}) is
\begin{equation}
	\chi_k(\eta) = \begin{cases}
		\chi_k^{\rm BD}(\eta) & (\eta<-H_{\rm inf}^{-1}) \\
		\alpha_k \chi_k^{\rm K}(\eta) + \beta_k \chi_k^{\rm K \ast}(\eta) & (\eta>-H_{\rm inf}^{-1})
	\end{cases}, \label{invac}
\end{equation}
where $\alpha_k$ and $\beta_k$ are the Bogoliubov coefficients. In order to determine them, we impose a junction condition that $\chi_k^{\rm in}(\eta)$ is continuous up to the first-order derivative with respect to $\eta$ at $\eta = -H_{\rm inf}^{-1}$.

\subsection{Gravitational particle creation} \label{subsec:acgpc}
\begin{figure}[tbp]
\centering
\includegraphics[width=.72\textwidth]{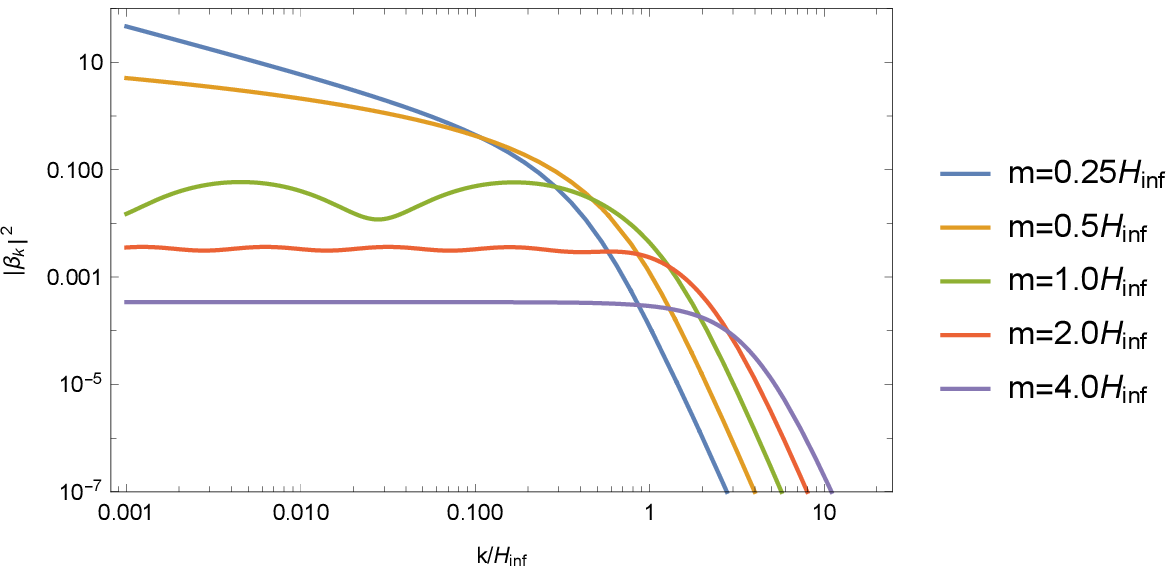} \\ \bigskip
\includegraphics[width=.72\textwidth]{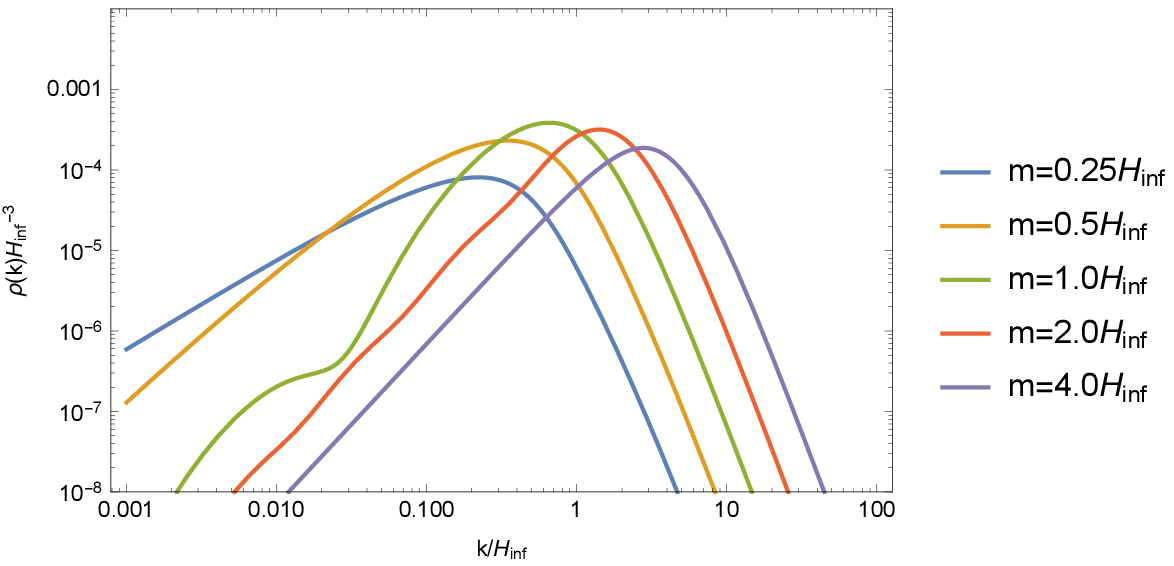}
\caption{\label{fig:ank} The power spectra of the Bogoliubov coefficient $|\beta_k|^2$ (upper panel) and those of  energy per the Hubble volume $\rho(k)H_{\rm inf}^{-3}$ (lower panel) of produced particles with various masses $0.25H_{\rm inf} \leq m \leq 4.0H_{\rm inf}$. For $k<m$, $|\beta_k|^2$ oscillates and its amplitude decreases as $m$ increases. (For $m=4.0H_{\rm inf}$, although $|\beta_k|^2$ seems to be constant for $k<m$, it does oscillate.) Once $k$ exceeds $m$, the spectra shows power law suppression. The power spectrum of the energy density has a peak around $k \sim m$.}
\end{figure}

The produced particle number density is given by the Bogoliubov coefficient $\beta_k$ in (\ref{invac}). After some calculations, we obtain
\begin{align}
	|\beta_k| = & \left| \frac{W[\chi_k^{\rm K}, \chi_k^{\rm BD}]}{W[\chi_k^{\rm K}, \chi_k^{\rm K \ast}]} \right|_{\eta=-H_{\rm inf}^{-1}} \nonumber \\
	= & \left| e^{\frac{1}{2}i\nu\pi} \right| \left| \sqrt{\frac{\pi}{6}} \:(2m^2H_{\rm inf})^{-1/6} x_f^{1/2} H^{(2)}_{1/3}\left(\frac{2}{3} x_f^{3/2}\right) \right. \nonumber \\
	& \times \left\{ \frac{\sqrt{\pi H_{\rm inf}}}{4} H^{(1)}_\nu\left(\frac{k}{H_{\rm inf}}\right) + \frac{1}{4}k\sqrt{\frac{\pi}{H_{\rm inf}}} \left[ H^{(1)}_{\nu-1}\left(\frac{k}{H_{\rm inf}}\right) - H^{(1)}_{\nu+1}\left(\frac{k}{H_{\rm inf}}\right) \right] \right\} \nonumber \\
	& + \sqrt{\frac{\pi}{6}} \:(2m^2H_{\rm inf})^{1/6} \left\{ \frac{1}{2} x_f^{-1/2} H^{(2)}_{1/3}\left(\frac{2}{3}x_f^{3/2}\right) \right. \nonumber \\
	& \left. \left. + \frac{1}{4} x_f \left[ H^{(2)}_{-2/3}\left(\frac{2}{3}x_f^{3/2}\right) - H^{(2)}_{4/3}\left(\frac{2}{3}x_f^{3/2}\right)  \right] \right\} \sqrt{\frac{\pi}{H_{\rm inf}}}H^{(1)}_\nu\left(\frac{k}{H_{\rm inf}}\right) \right| , \label{betaa}
\end{align}
where $x_f = \frac{k^2+m^2}{(2m^2H_{\rm inf})^{2/3}}$. The power spectrum of the number density $n_k=|\beta_k|^2$ and that of the energy density $\rho_k$ with various $m$ are shown in Figure \ref{fig:ank}.
If $m/H_{\rm inf}$ is fixed, produced energy in the Hubble volume $\rho_k H_{\rm inf}^{-3}$ does not depend on $H_{\rm inf}$. The power spectrum of the number density oscillates when $k<m$ and decreases when $k>m$.

The energy density has a peak around $k \sim m$. It has been widely claimed that particle creation is exponentially suppressed in the UV (adiabatic) limit \cite{Chung2001,Kuzmin1999}, however, an asymptotic form of the Bogoliubov coefficient (\ref{betaa}) in the high $k$ limit is
\begin{equation}
	|\beta_k|^2 \simeq \left( \frac{3}{8} m^2H_{\rm inf}^2 k^{-4} \right)^2, \label{UVb}
\end{equation}
which only shows power law suppression. This $m$-dependence $|\beta_k|^2 \propto m^4$ is consistent with the former studies using the Born \cite{Ford1987} or the WKB approximation \cite{Mamaev1976}. The dependence on $H_{\rm inf}$ and $k$ significantly changes according to the evolution law of the scale factor, so that these former results cannot apply to the present case, since \cite{Ford1987} assumes that $m^2 (a^2(-\infty) - a^2(+\infty)) = 0$ and \cite{Mamaev1976} assumes an existence of a singularity.

\begin{figure}[tbp]
\centering
\includegraphics[width=.65\textwidth]{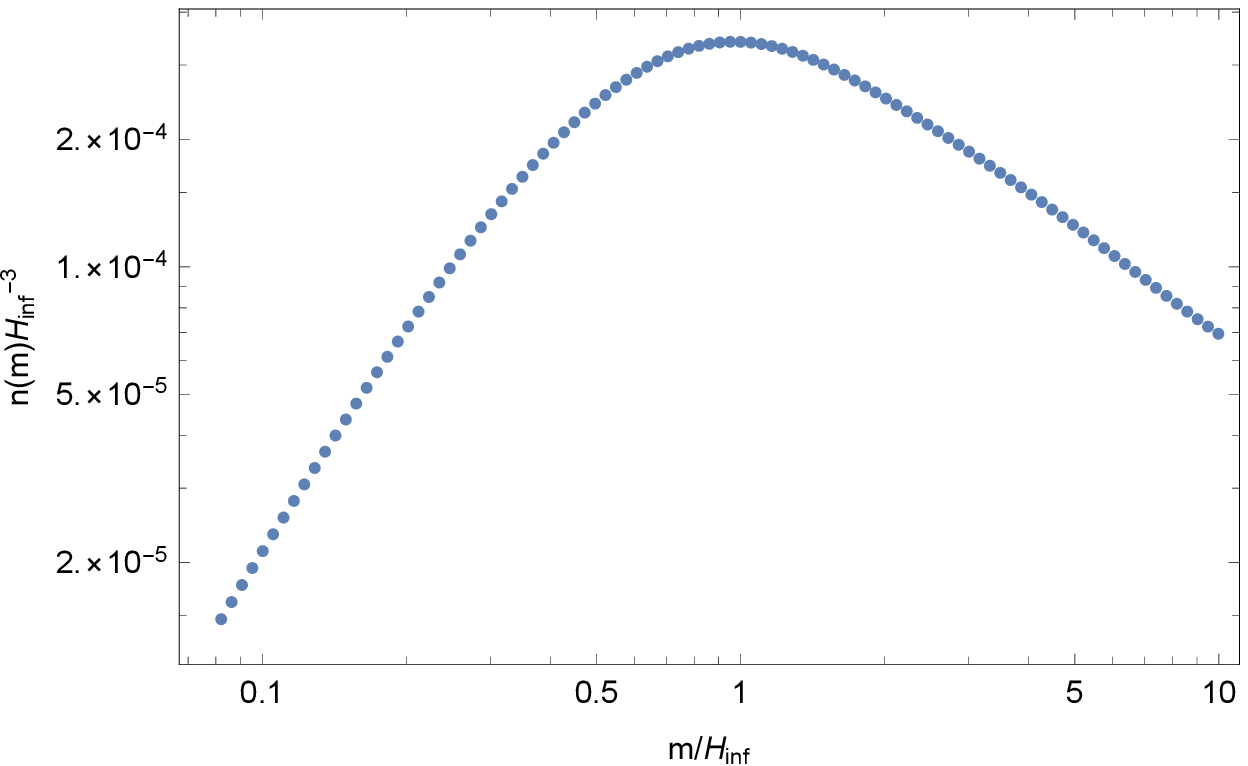} \\ \bigskip
\includegraphics[width=.65\textwidth]{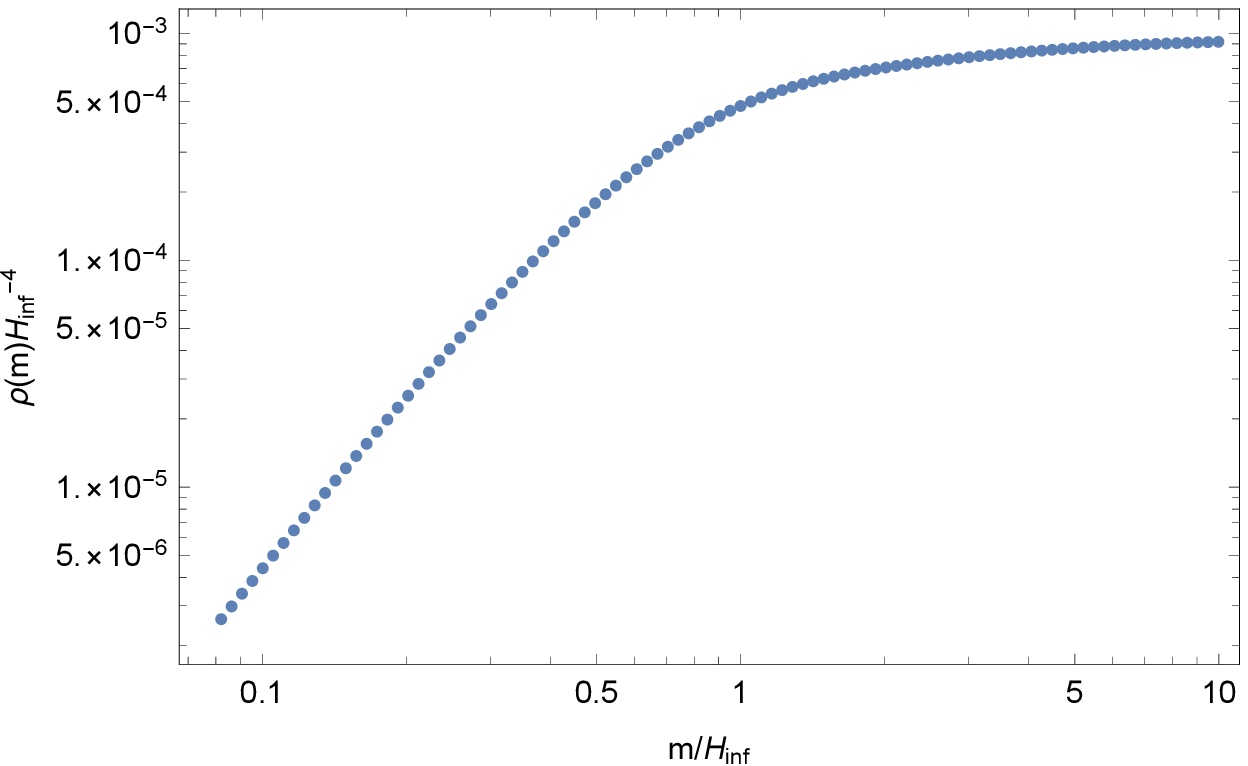}
\caption{\label{fig:anm} The number (upper panel) and the energy density (lower panel) of produced particle with various masses $0.08H_{\rm inf} < m < 10.0H_{\rm inf}$. They are scaled by $H_{\rm inf}^{-3}$ and $H_{\rm inf}^{-4}$, respectively. Even particles much heavier than $H_{\rm inf}$ are abundantly produced.}
\end{figure}

The number and the energy density of produced particle are shown in Figure \ref{fig:anm}.
We obtain
\begin{equation}
	\rho_\phi \sim \begin{cases}
		1.4 \times 10^{-3} m^2 H_{\rm inf}^2 & (m<H_{\rm inf}) \\
		10^{-3}H_{\rm inf}^4 & (H_{\rm inf}<m)
	\end{cases} \label{rhoan}
\end{equation}
by fitting. The energy density of the scalar field seems to approach a non-vanishing value $\rho_\phi \sim 10^{-3}H_{\rm inf}^4$ as the mass increases. This appears to imply that superheavy particles could be produced much more than we expected. However, we should not take it literally since we have adopted a rather unnatural scale factor of which second-order derivative is discontinuous at the transition epoch. A realistic scale factor must be always smooth. This case is considered in section \ref{sec:nc}.

\section{Numerical calculation} \label{sec:nc}
\subsection{Smoothly connected scale factor} \label{subsec:ncsf}
It is known that if a particle is non-conformally coupled to gravity and the second-order derivative of the scale factor is discontinuous, then $\rho_\phi$ diverges due to a $\delta$ function-like behavior of the derivative of the scalar curvature \cite{Ford1987}. It may be worth calculating the particle production rate with a scale factor of class $C^\infty$ although the particle is conformally coupled to gravity in our case and hence the derivative of the scalar curvature does not emerge explicitly.

\begin{figure}[tbp]
\centering
\includegraphics[width=.55\textwidth]{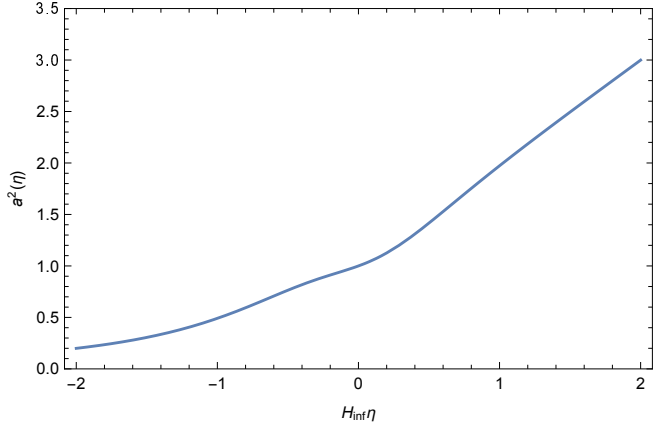}
\caption{\label{fig:scale} A scale factor of class $C^\infty$ which connects the inflation stage and the kination stage with a finite transition time scale $\Delta\eta$. The squared scale factor $a^2$ smoothly changes its dependence on the conformal time from $\eta^{-2}$ to $\eta$. We set $\Delta\eta = 0.5 H_{\rm inf}^{-1}$ in this graph.}
\end{figure}

In realistic models, the scale factor is derived from a background dynamics of inflation, and then the detail of the scale factor differs from model to model. Instead of focusing on a specific model, we approximate the scale factor with a transition from inflation to kination stages whose time scale is $\Delta\eta>0$ by a function such as
\begin{equation}
	a^2(\eta) = \frac{1}{2} \left[ \left(1 - \tanh\frac{\eta}{\Delta\eta}\right)\frac{1}{1 + H_{\rm inf}^2 \eta^2} + \left(1 + \tanh\frac{\eta}{\Delta\eta}\right)(1 + H_{\rm inf} \eta) \right], \label{scalen}
\end{equation}
which is shown in Figure \ref{fig:scale}. $a^2(\eta)$ monotonically increases when $\Delta\eta<1.797H_{\rm inf}^{-1}$ and remains positive for all $\eta$ when $\Delta\eta<1.991H_{\rm inf}^{-1}$. If $\Delta\eta$ becomes much larger than $H_{\rm inf}$, then the scale factor gets distorted. With $\Delta\eta \sim H_{\rm inf}$, this scale factor actually agrees with that of the $k$-inflation \cite{Picon1999} and the kinetically driven G-inflation \cite{Kobayashi2010} within $15\%$ in the transition regime. The values of $\Delta\eta$ are $\Delta\eta = 1.25H_{\rm inf}$ for the $k$-inflation described in \cite{Picon1999}, and $\Delta\eta = 1.41H_{\rm inf}$ for the kinetically driven G-inflation whose Lagrangian is
\begin{equation}
	\mathcal{L}_\varphi = \tanh\left( \frac{\varphi}{M_G} \right) X + \frac{X^2}{2M^4},
\end{equation}
where $\varphi$ denotes the inflaton, $X=-\frac{1}{2}(\nabla\varphi)^2$ and $M=10^{-4}M_G$.

\subsection{Numerical results and effective temperature} \label{subsec:ncnr}
\begin{figure}[tbp]
\centering
\includegraphics[width=.74\textwidth]{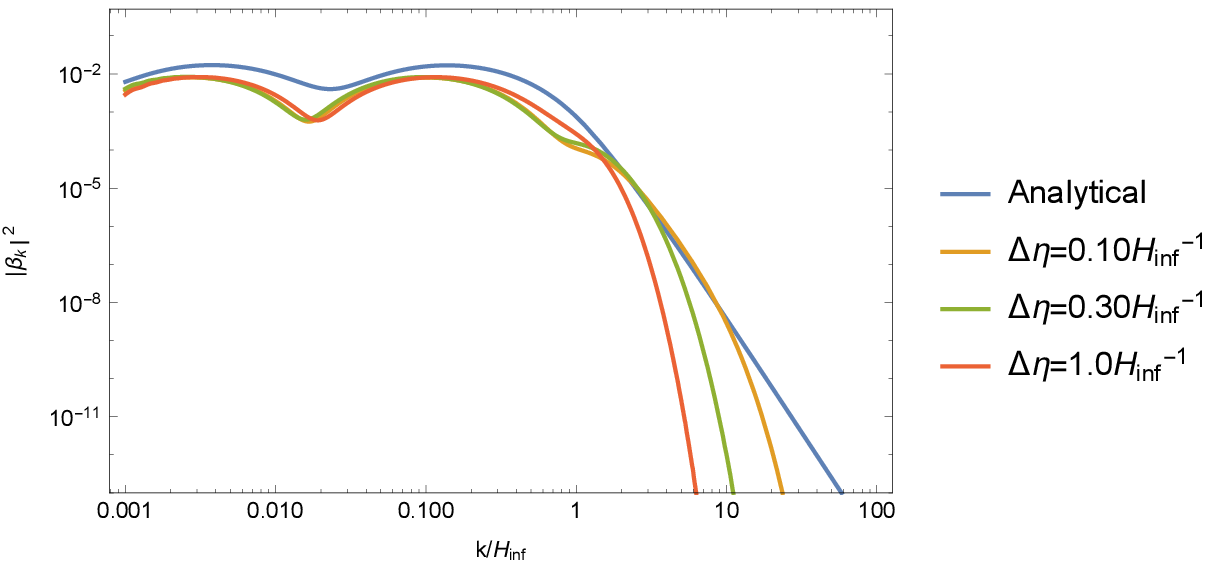} \\ \bigskip
\includegraphics[width=.74\textwidth]{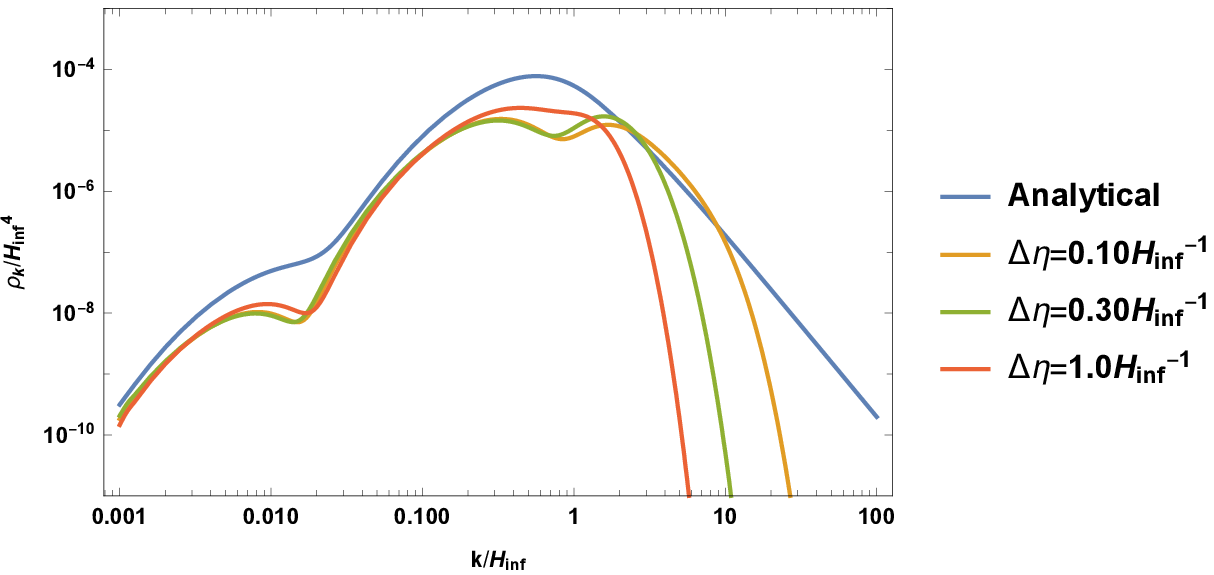}
\caption{\label{fig:nmc} The power spectra of the Bogoliubov coefficient $|\beta_k|^2$ (upper panel) and those of  energy $\rho_k$ (lower panel) of produced particles with the mass $m=1.0H_{\rm inf}$ and various transition time scales $0.10 H_{\rm inf}^{-1} \leq \Delta\eta \leq 1.0 H_{\rm inf}^{-1}$. The difference between the analytical and the numerical solutions in the IR region comes from the difference in scale factor. Once $k$ exceeds $\Delta\eta^{-1}$, they are suppressed exponentially.}
\end{figure}

\begin{figure}[tbp]
\centering
\includegraphics[width=.74\textwidth]{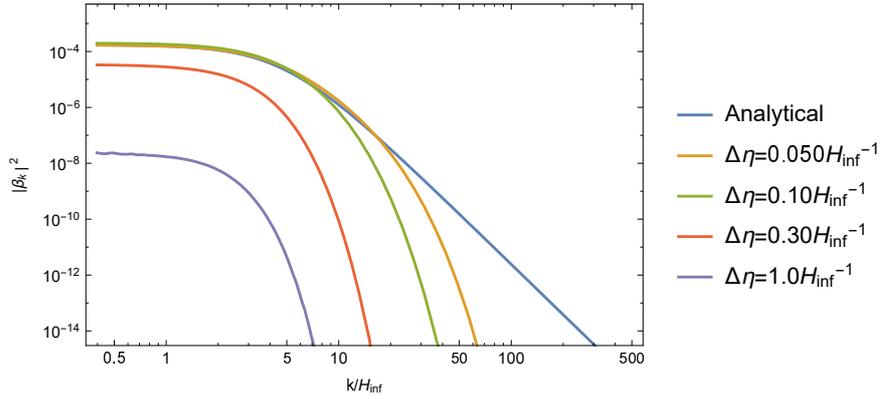}
\caption{\label{fig:nmc5} The power spectra of the Bogoliubov coefficient $|\beta_k|^2$ with the mass $m=5.0H_{\rm inf}$ and various transition time scales $0.10 H_{\rm inf}^{-1} \leq \Delta\eta \leq 1.0 H_{\rm inf}^{-1}$. The numerically calculated Bogoliubov coefficients asymptotically approach the analytical one as $\Delta\eta \to 0$. Once $k$ exceeds $\Delta\eta^{-1}$, $|\beta_k|^2$ is exponentially suppressed and the height of the plateau in $k<m$ is also suppressed if $m>\Delta\eta^{-1}$.}
\end{figure}

Although (\ref{scalen}) is more natural than the previous one (\ref{scale}), we cannot analytically solve the equation of motion (\ref{eom}) with this scale factor. Thus, we numerically solve (\ref{eom}) assuming that the mode function is the same as the positive frequency mode in the Minkowski space at far past $\eta=\eta_{\rm min} \ll -k^{-1} \; ({\rm i.e.\ } |k\eta|\gg1)$ and calculate the Bogoliubov coefficients compared with the k-vacuum at far future $\eta=\eta_{\rm max} \gg H_{\rm inf}^{-1} \; ({\rm i.e.\ } x\gg1)$. The obtained power spectra of the number density are shown in Figure \ref{fig:nmc} and \ref{fig:nmc5} for $m=1.0H_{\rm inf}$ and $m=5.0H_{\rm inf}$, respectively.

When $k$ is smaller than $\Delta\eta^{-1}$, $|\beta_k|^2$ shows the same general behavior as the analytical solution. This is quite natural because a typical time scale of a particle with a momentum $k$ is $k^{-1}$ and then the particle cannot feel the smoothness of the scale factor (\ref{scalen}) when $k^{-1}>\Delta\eta$. The difference between the analytical and the numerical solutions in the IR region when $m=1.0H_{\rm inf}$ (Figure \ref{fig:nmc}) comes from the difference of the scale factor. The scale factor (\ref{scalen}) is made by combining $\frac{1}{1 + H_{\rm inf}^2 \eta^2}$ with $1 + H_{\rm inf} \eta$. However, since we cannot analytically solve (\ref{eom}) with the scale factor $\frac{1}{1 + H_{\rm inf}^2 \eta^2}$, we could not help but use a slightly different scale factor in the analytical calculation. On the other hand, $|\beta_k|^2$ asymptotically approaches nothing but the analytical one in the IR region as $\Delta\eta$ decreases when $m=5.0H_{\rm inf}$ (Figure \ref{fig:nmc5}). It is explained as follows. Writing (\ref{eom}) in terms of the physical time $t$, we obtain
\begin{equation}
	\ddot{\chi_k} + H\dot{\chi_k} + \left[ \left(\frac{k}{a}\right)^2 + m^2 \right]\chi_k = 0,
\end{equation}
where the dot and $H \equiv \dot{a}/a$ denote $t$ derivative and the Hubble parameter, respectively, and $k/a$ is the physical wave number. As can be seen here, the difference in $a$ is negligible if $m$ is sufficiently large.

A significant difference appears in the UV region. When $k$ is larger than $\Delta\eta^{-1}$ and the particle can feel the smoothness or, in other words, the adiabatic nature of the scale factor, $|\beta_k|^2$ is suppressed not as a power-law but exponentially and the height of the plateau in $k<m$ is also suppressed if $m>\Delta\eta^{-1}$. This result is consistent with \cite{Chung1999} in which they claim that $n_\phi \sim (m/H_{\rm inf})^{-(2s-3)}$ when $(d^q a/d\eta^q) / a^{q+1}$ has a discontinuity of the first kind if and only if $q \geq s$, and $n_\phi$ is suppressed exponentially when the scale factor is of class $C^\infty$.

\begin{figure}[tbp]
\centering
\includegraphics[width=.78\textwidth]{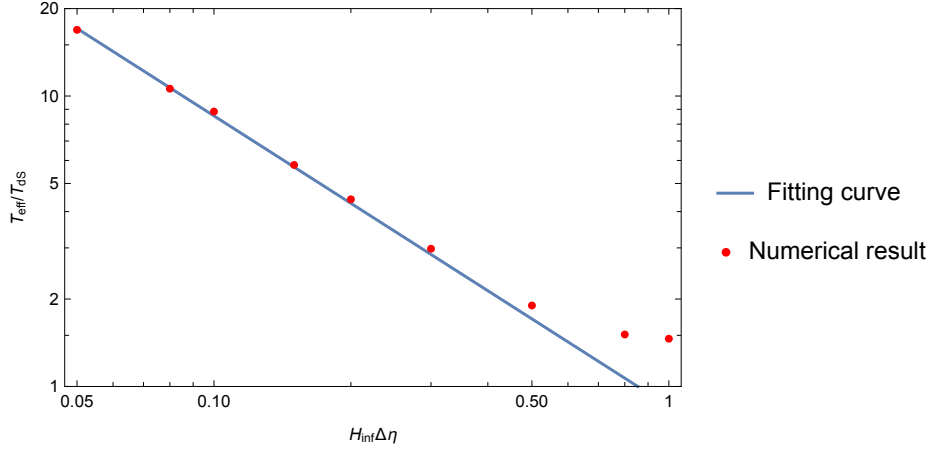}
\caption{\label{fig:temp_d} The dependence of the effective temperature just after inflation ends (normalized by the Hawking temperature of de Sitter space $T_{\rm dS}=H_{\rm inf}/2\pi$) on the transition time scale. The red dots and the blue line represent the numerical results and the fitting curve $T_{\rm eff}=0.85T_{\rm dS}(H_{\rm inf}\Delta\eta)^{-1}$, respectively.}
\end{figure}

We also find that this exponential suppression in the UV region is well approximated by the Boltzmann factor $e^{-E/T_{\rm eff}}$ with $E\equiv\sqrt{k^2+m^2}$. Note, however, that this does not mean that the number density is given by the thermal distribution of a massive particle at temperature $T_{\rm eff}$, which is much larger than the actual value (see Section \ref{subsec:ncd}). The effective temperature $T_{\rm eff}$ at the time of particle creation is independent of $m$ and depends on $H_{\rm inf}$ and $\Delta\eta$ as shown in Figure \ref{fig:temp_d}.
When $\Delta\eta$ is smaller than around $0.50H_{\rm inf}^{-1}$, the effective temperature is inversely proportional to $\Delta\eta$ and the fitting curve is $T_{\rm eff}=0.85T_{\rm dS}(H_{\rm inf}\Delta\eta)^{-1}$. When $\Delta\eta$ is larger than around $0.50H_{\rm inf}^{-1}$, the effective temperature deviates from the fitting curve. This is because of the distortion of the scale factor (\ref{scalen}) when $\Delta\eta>H_{\rm inf}$.

\subsection{Reheating temperature} \label{subsec:ncd}
In the previous subsubsection, we have derived the effective temperature just after inflation (at the beginning of the kination stage) from the power spectrum of the Bogoliubov coefficient. On the other hand, we can calculate the ``real'' reheating temperature provided that gravitational particle creation is a dominant reheating mechanism and produced particles decay into radiation. Since we have obtained the power spectrum of the Bogoliubov coefficient and that of the energy density, we can calculate the number density $n_\phi$ and the energy density $\rho_\phi$ of the scalar field by integration,
\begin{align}
	n_\phi &= \int^\infty_0 \frac{4\pi k^2 dk}{(2\pi)^3} |\beta_k|^2 \\
	\rho_\phi &= \int^\infty_0 \frac{4\pi k^2 dk}{(2\pi)^3} \sqrt{m^2 + k^2} |\beta_k|^2. \label{rhoint}
\end{align}
We assume that the scalar field decays all at once and consider following two cases. In the first case the scalar particles decay immediately after their creation, and in the second case they decay when its energy density exceeds that of the inflaton. The former is for the sake of comparison with previous analyses \cite{Ford1987,Kunimitsu2012,Nishi2016}, while the latter gives the maximum possible reheating temperature in the model.

\begin{figure}[tbp]
\centering
\includegraphics[width=.72\textwidth]{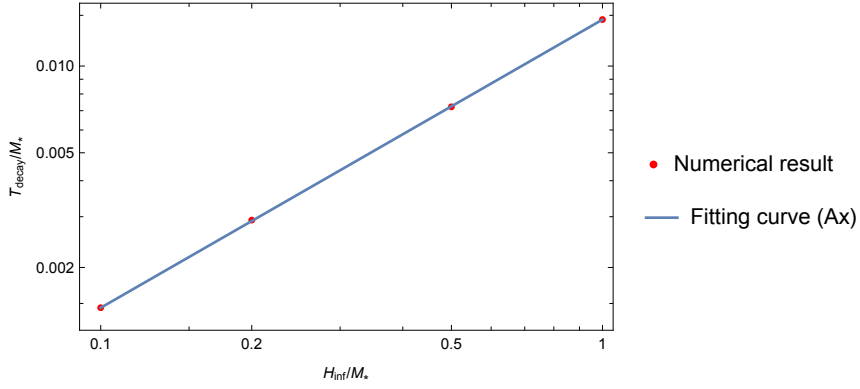}
\caption{\label{fig:THmd} Case A. The dependence of $T_{\rm decay}$ on the Hubble parameter during inflation when $\Delta\eta=0.5H_{\rm inf}^{-1}$ and $m=2.0H_{\rm inf}$. The red dots and the blue line represent the numerical results and the graph of $T=0.014H_{\rm inf}$, respectively. $T_{\rm decay}$ is proportional to $H_{\rm inf}$ as expected.}
\end{figure}

\begin{figure}[tbp]
\centering
\includegraphics[width=.72\textwidth]{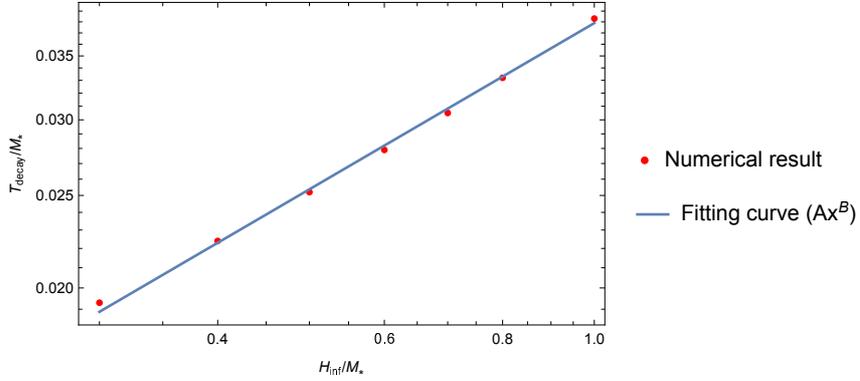}
\caption{\label{fig:TH} Case B. The dependence of $T_{\rm decay}$ on the Hubble parameter during inflation when $\Delta\eta=0.5M_\ast^{-1}$ and $m=2.0M_\ast$. The red dots and the blue line represent the numerical results and the graph of $T=0.038 (H_{\rm inf}/M_\ast)^{0.58}M_\ast$, respectively.}
\end{figure}

\begin{figure}[tbp]
\centering
\includegraphics[width=.72\textwidth]{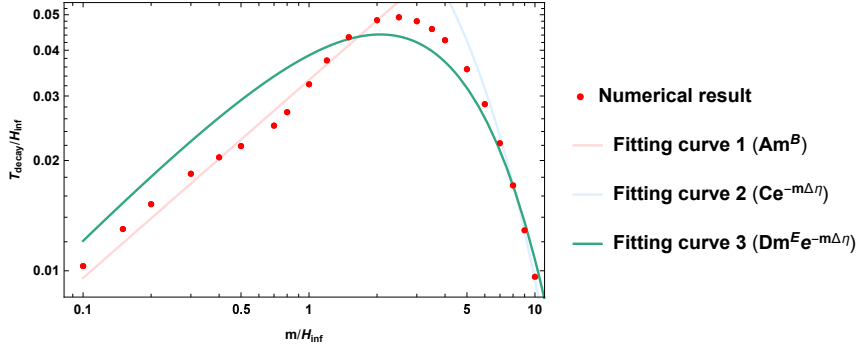}
\caption{\label{fig:Tm} Case C. The dependence of $T_{\rm decay}$ on the mass of the scalar field when $\Delta\eta=0.30H_{\rm inf}^{-1}$. The red dots, the light red line the light blue line and the green line represent the numerical results, the graph of $T=0.033 (m/H_{\rm inf})^{0.54}H_{\rm inf}$, the graph of $T=0.19 e^{-m\Delta\eta} H_{\rm inf}$ and $T=a(m/H_{\rm inf})^b e^{-m\Delta\eta}H_{\rm inf}$, where $a=0.052$ and $b=0.62$, respectively.}
\end{figure}

\begin{figure}[tbp]
\centering
\includegraphics[width=.80\textwidth]{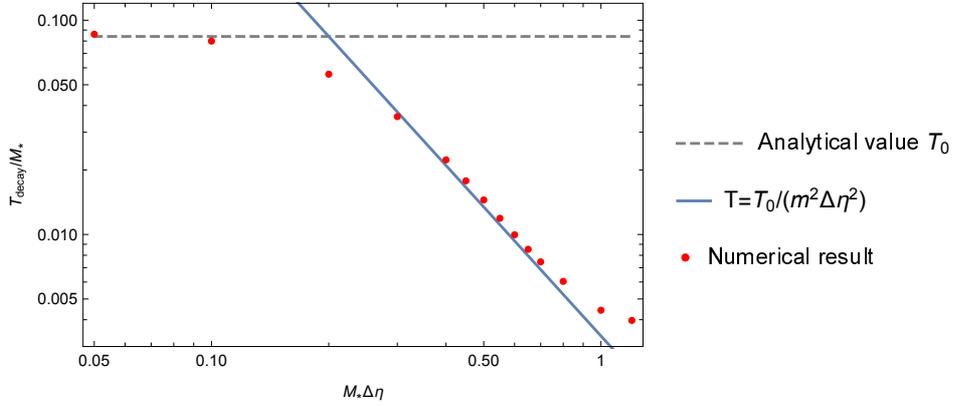}
\caption{\label{fig:TRHm5} Case D. The dependence of $T_{\rm decay}$ on the transition time scale $\Delta\eta$ when $H_{\rm inf}=1.0M_\ast$ and $m=5.0M_\ast$. The red dots, the gray dashed line and the blue line represent the numerical results, the temperature $T_0$ derived from (\ref{betaa}) and the graph of $T=T_0/(m^2\Delta\eta^2)$, respectively.}
\end{figure}

First, the case that the scalar particles decay and thermalize instantly. The produced radiation energy density $\rho_r$ is equal to $\rho_\phi$ given in Figure \ref{fig:anm} and we define the temperature at that time by
\begin{equation}
	\rho_\phi=\rho_r = \frac{\pi^2}{30}g_\ast T_{\rm decay}^4, \label{rhoT}
\end{equation}
where we take $g_\ast = 106.75$. The temperature $T_{\rm decay}$ depends on all of $m$, $H_{\rm inf}$ and $\Delta\eta$. We analyze the dependence on them one by one as follows.
\begin{enumerate}
\renewcommand{\labelenumi}{\Alph{enumi}.}
	\item vary $H_{\rm inf}$, ($\Delta\eta$, $m$) while fixing $H_{\rm inf}\Delta\eta$ and $m/H_{\rm inf}$.
	\item vary $H_{\rm inf}$ while fixing $\Delta\eta$ and $m$.
	\item vary $m$ while fixing $\Delta\eta$ and $H_{\rm inf}$.
	\item vary $\Delta\eta$ while fixing $m$ and $H_{\rm inf}$.
\end{enumerate}
\begin{description}
\item[Case A.] The result for $\Delta\eta=0.5H_{\rm inf}^{-1}$ and $m=2.0H_{\rm inf}$ with various Hubble parameters $0.1M_\ast \leq H_{\rm inf} \leq 1.0M_\ast$, where $M_\ast$ is an arbitrary mass scale, is shown in Figure \ref{fig:THmd}. $T_{\rm decay}$ is exactly proportional to $H_{\rm inf}$ within the numerical error as expected.
\item[Case B.] The result for $\Delta\eta=0.5M_\ast^{-1}$ and $m=2.0M_\ast$ with various Hubble parameters $0.3M_\ast \leq H_{\rm inf} \leq 1.0M_\ast$ is shown in Figure \ref{fig:TH}. $T_{\rm decay}$ seems proportional to $H_{\rm inf}^{0.58}$.
\item[Case C.] The result for $H_{\rm inf}=1.0M_\ast$ and $\Delta\eta=0.3M_\ast^{-1}$ with various masses $0.10 M_\ast \leq m \leq 10.0 M_\ast$ is shown in Figure \ref{fig:Tm}. $T_{\rm decay}$ takes the maximum at $m \sim \Delta\eta^{-1}$.When $m$ is smaller than $\Delta\eta^{-1}$, $T_{\rm decay}$ increases proportionally to $m^{0.54}$ while oscillating. This index is almost the same as that of the Hubble parameter $0.58$. Therefore, we find that $m$ and $H_{\rm inf}$ have approximately the same contribution to $T_{\rm decay}$ and so $T_{\rm decay} \propto \sqrt{mH_{\rm inf}}\ (\therefore\; \rho_\phi \propto m^2H_{\rm inf}^2)$. Once $m$ exceeds $\Delta\eta^{-1}$, $T_{\rm decay}$ is exponentially suppressed as $T_{\rm decay} \sim e^{-m\Delta\eta}$. This behavior is just like the power spectrum of the energy density (Figure \ref{fig:nmc}). As a whole, the graph of $T_{\rm decay}$ can be well fitted by
\begin{equation}
T=a\left(\frac{m}{H_{\rm inf}}\right)^b e^{-m\Delta\eta}H_{\rm inf},
\end{equation}
where $a=0.052$ and $b=0.62$.
\item[Case D.] The result for $H_{\rm inf}=1.0M_\ast$ and $m=5.0M_\ast$ with various transition time scales $0.050 M_\ast^{-1} \leq \Delta\eta \leq 1.0 M_\ast^{-1}$ is shown in Figure \ref{fig:TRHm5}. $T_{\rm decay}$ asymptotically approaches $T_0$ at $\Delta\eta \to 0$, where $T_0$ refers to the radiation temperature calculated in the same way as $T_{\rm deacy}$ in (\ref{rhoT}) for the analytic model shown in Figure \ref{fig:anm}. Once $\Delta\eta$ exceeds $m^{-1}$, $T_{\rm decay}$ decreases by $\Delta\eta^{-2}$. Although it seems that $T_{\rm decay}$ goes away from the line of $T=T_0/m^2\Delta\eta^2$, this comes from the distortion of the scale factor (\ref{scalen}) when $\Delta\eta>H_{\rm inf}$.
\end{description}
The results of cases A-C can be summarized as
\begin{equation}
	T_{\rm decay} \simeq 5 \times 10^{11} e^{-m\Delta t} \left(\frac{m}{10^{13}{\rm GeV}}\right)^{1/2} \left(\frac{H_{\rm inf}}{10^{13}{\rm GeV}}\right)^{1/2} \; {\rm GeV} \label{Tdecay}
\end{equation}
or in terms of the energy density,
\begin{equation}
	\rho_\phi \simeq 2 \times 10^{-4} e^{-4m\Delta t} m^2 H_{\rm inf}^2, \label{rhopro}
\end{equation}
where we have ignored the differences of the coefficients of factors smaller than $1.5$ and replace $\Delta\eta$ with $\Delta t$ since the scale factor at the end of inflation is unity. In contrast to the previous analytical calculation (Figure \ref{fig:anm}), the produced energy density does not level off at a certain value. This is because the $\Delta\eta \to 0$ limit of the scale factor (\ref{scalen}) has a discontinuity in terms of the first-order time derivative, and then there is no limitation. The radiation energy density at time $t$ after the decay of the scalar field is
\begin{equation}
	\rho_r (t) = \frac{\pi^2}{30}g_\ast T_{\rm decay}^4 a(t)^{-4}, \label{rhorad}
\end{equation}
where we let the scale factor at the beginning of the kination stage be unity just like in section \ref{subsec:acmf}. On the other hand, the background energy density after inflation, now dominated by the kinetic energy density of the infaton $\rho_I$, which evolves as
\begin{equation}
	\rho_I (t) = 3M_G^2 H_{\rm inf}^2 a(t)^{-6}. \label{rhoIt}
\end{equation}
It follows from (\ref{Tdecay}), (\ref{rhorad}) and (\ref{rhoIt}) and the reheating condition $\rho_r = \rho_I$ that the reheating temperature is given by
\begin{equation}
	T_{\rm RH} \simeq 2\times10^4 e^{-3m\Delta t} \left(\frac{m}{10^{13}{\rm GeV}}\right)^{3/2} \left(\frac{H_{\rm inf}}{10^{13}{\rm GeV}}\right)^{1/2} {\rm GeV}. \label{TRH1}
\end{equation}
This is smaller than the case reheating proceeds through production of minimally coupled massless scalar particles \cite{Kunimitsu2012}, which predicts $T_{\rm RH} \simeq 7 \times 10^5\:$GeV for $H_{\rm inf}=10^{13}\:$GeV. This is because the time dependence of the effective mass term (i.e. the last term of (\ref{eom})) is much milder in the present model.

Next, the case that the scalar particles decay when their energy density exceeds that of the inflaton. The number density of the scalar field $n_\phi$ decreases as $a^{-3}$. Since the power spectrum has its peak at $k \sim m$ (Figure \ref{fig:nmc},\ref{fig:nmc5}) at the beginning of the kination stage, the scalar field soon becomes non-relativistic as the universe expands. Therefore, the energy density of the scalar field $\rho_\phi$ is
\begin{equation}
	\rho_\phi (t) \simeq m n_\phi (t_\ast) a(t)^{-3}, \label{rhophit}
\end{equation}
where $t_\ast$ is the time at the beginning of the kination stage. It follows from (\ref{rhoIt}), (\ref{rhophit}) and the reheating condition $\rho_\phi = \rho_I$ that
\begin{equation}
	\rho_r (t_{\rm RH}) = \frac{m^2 n_\phi (t_\ast)^2}{3M_G^2 H_{\rm inf}^2}.
\end{equation}
By substituting (\ref{rhopro}) we obtain
\begin{equation}
	T_{\rm RH} \simeq 7\times10^7 e^{-2m\Delta t} \left(\frac{m}{10^{13}{\rm GeV}}\right) \left(\frac{H_{\rm inf}}{10^{13}{\rm GeV}}\right)^{1/2} {\rm GeV}, \label{TRH2}
\end{equation}
which gives the maximum possible reheating temperature for given value of $m$, $H_{\rm inf}$ and $\Delta t$.

Lastly we comment on the transition time scale in ordinary inflationary models. In most cases, $\Delta t$ is the order of $H_{\rm inf}^{-1}$ since it is the only natural time scale in the single-field inflationary universe including most of kinetically driven inflations \cite{Picon1999,Kobayashi2010} or some potential-driven inflations \cite{Peebles1999}. However, a very steep potential or a multi-field model may realize a small $\Delta t$. The way to superheavy particles is not completely closed.

\section{Conclusion} \label{sec:c}
In this paper, we analyzed the gravitational particle creation of the massive scalar particle of which mass may be larger than the Hubble parameter during inflation without the perturbative expansion. When the inflation and the kination stages are connected continuously only up to the first-order derivative, the produced energy density asymptotically approaches a non-vanishing value as the mass increases. When the inflation and the kination stages are combined into the one analytical scale factor of class $C^\infty$ with a finite transition time scale $\Delta\eta$, which is almost identical to the physical time scale $\Delta t$ as we are taking $a=1$ at the transition epoch, our numerical calculation shows that the produced amount of a particle heavier than $\Delta t^{-1}$ is exponentially suppressed as shown in (\ref{rhopro}). The representation of reheating temperature is obtained as (\ref{TRH1}) and (\ref{TRH2}) in the case that the scalar field decays instantly and in the case that the scalar field decays when its energy density exceeds that of the inflaton, respectively. We find a somewhat lower reheating temperature (\ref{TRH1}) than in \cite{Kunimitsu2012} if the produced particles decay immediately. If they have a longer life time, on the other hand, the reheating temperature can be much higher, (\ref{TRH2}). We also note that production of massive particles discussed here may also be helpful to realize non-thermal baryo/leptogenesis.

Finally, although we considered only the case the mass of the produced particle is constant, it may be generated by an expectation value of the Higgs field, which may be spatially fluctuating due to the long-wave quantum fluctuations generated during inflation. Then this may yield a modulated reheating scenario. This issue will be reported elsewhere \cite{Hashiba}.

\acknowledgments
We thank Kohei Kamada and Alexei Alexandrovich Starobinsky for useful comments. Shijie Wang also gave SH advice on numerical calculations. SH was supported by the Advanced Leading Graduate Course for Photon Science (ALPS). The work of JY was supported by JSPS KAKENHI, Grant JP15H02082 and Grant on Innovative Areas JP15H05888.


\end{document}